\documentclass[fleqn,usenatbib]{mnras}
\usepackage{newtxtext,newtxmath}
\usepackage[T1]{fontenc}
\DeclareRobustCommand{\VAN}[3]{#2}
\let\VANthebibliography\thebibliography
\def\thebibliography{\DeclareRobustCommand{\VAN}[3]{##3}\VANthebibliography}
\usepackage{graphicx}	
\usepackage{color,units}
\usepackage{comment}
\usepackage{caption}
\usepackage{subcaption}
\usepackage{nicefrac}
\usepackage{ulem}

\newcommand{\order}[1]{$\mathcal{O}(#1)$}

\newcommand{\new}[1]{\textcolor{black} {#1}}

\title[From Gravitational Waves to Globular Clusters]{Gravitational Waves as a Probe of Globular Cluster Formation and Evolution}

\author[I. M. Romero-Shaw et al.]{
Isobel M. Romero-Shaw,$^{1,2}$\thanks{isobel.romero-shaw@monash.edu}
Kyle Kremer,$^{3,4}$
Paul D. Lasky,$^{1,2}$
Eric Thrane$^{1,2}$
and Johan Samsing$^{5}$
\\
$^{1}$Monash Centre for Astrophysics, School of Physics and Astronomy, Monash University, VIC 3800, Australia\\
$^{2}$OzGrav: The ARC Centre of Excellence for Gravitational-wave Discovery, Clayton, VIC 3800, Australia\\
$^{3}$TAPIR, California Institute of Technology, Pasadena, CA 91125, USA\\
$^{4}$The Observatories of the Carnegie Institution for Science, Pasadena, CA 91101, USA\\
$^{5}$Niels Bohr International Academy, The Niels Bohr Institute, Blegdamsvej 17, 2100 Copenhagen, Denmark\\
}

\date{Accepted XXX. Received YYY; in original form ZZZ}

\pubyear{2020}

\begin{document}
\label{firstpage}
\pagerange{\pageref{firstpage}--\pageref{lastpage}}
\maketitle

\begin{abstract}
Globular clusters are considered to be likely breeding grounds for compact binary mergers.
In this paper, we demonstrate how the gravitational-wave signals produced by compact object mergers can act as tracers of globular cluster formation and evolution.
Globular cluster formation is a long-standing mystery in astrophysics, with multiple competing theories describing when and how globular clusters formed.
The limited sensitivity of electromagnetic telescopes inhibits our ability to directly observe globular cluster formation.
However, with future audio-band detectors sensitive out to redshifts of $z \approx 50$ for GW150914-like signals, gravitational-wave astronomy will enable us to probe the Universe when the first globular clusters formed.
We simulate a \new{population} of \new{binary black hole mergers} from theoretically-motivated globular cluster formation models, \new{and construct redshift measurements consistent with the predicted accuracy of third-generation detectors}.
\new{
We show that we can locate the peak time of a cluster formation epoch during reionisation to within \unit[0.05]{Gyr} after one year of observations.
The peak of a formation epoch that coincides with the Universal star formation rate can be measured to within \unit[0.4]{Gyr}---\unit[10.5]{Gyr} after one year of observations, depending on the relative weighting of the model components.} 
\end{abstract}




\section{Introduction}
The first detections of gravitational waves, made over the last five years~\citep{GWTC-1, GWTC-2}, provide a new lens through which to observe the Universe. 
Advanced LIGO~\citep{LIGO} and Virgo~\citep{virgo} have confirmed the existence of multiple phenomena that, prior to the era of gravitational-wave astronomy, had only been theoretically proposed; stellar-mass binary black holes (BBH)~\citep{abbott16_01BBH}, merging neutron stars~\citep{abbott17_gw170817_detection, abbott17_gw170817_multimessenger}, and intermediate-mass black holes~\citep{GW190521_disco} have all been directly observed with gravitational waves.

We are accruing gravitational-wave observations of merging black holes at an accelerating rate~\citep{abbott16_01BBH, GWTC-1, GWTC-2,GWTC-2_RnP}. 
This abundance of BBH merger detections presents a variety of puzzles across theoretical astrophysics.
One such question is how BBH systems that merge within the age of the Universe are assembled. 
If the binary evolves in isolation, this outcome may be achieved via the common envelope process~\citep[see, e.g.,][]{Livio88, Bethe98, Ivanova13, Kruckow16}, stable mass transfer of a stellar secondary onto the primary black hole~\citep{Heuvel17,Neijssel19,Bavera21}, chemically homogeneous evolution~\citep{deMink10, deMink16} and/or ambient gas-driven fallback~\citep{Tagawa18}.
Alternatively, the compact object binary may form dynamically. 
In this case, the two components evolve separately, only encountering one another once they are already black holes. 
For this encounter to take place, the components must reside in an environment facilitating dynamical interactions.
Such environments include active galactic nuclei~\citep[e.g.,][]{Grobner20}, nuclear star clusters~\citep[e.g.,][]{Hoang2018, Fragione18}, young massive clusters~\citep{DiCarlo19} and globular clusters~\citep[e.g.,][]{Rodriguez15, Hong18}. 
In globular clusters, mass segregation leads to the formation of a dark compact-object core~\citep[see, e.g.,][]{Morscher15, DRAGON, Kyle_data_paper}, where black holes may interact and merge dynamically. 
\new{Evidence from LIGO--Virgo's third observing run suggests that a substantial fraction ($\approx25-93\%$ with $90\%$ credibility) of merging BBH form dynamically~\citep{GWTC-2_RnP}; see also \cite{GW190521_disco, GW190521_implications, RoShoGW190521, Zevin2020}.}

The gravitational-wave signal from a binary compact object merger carries information about the source's component masses, component spins, and orbital eccentricity. 
These parameters can be used to distinguish which formation channel the binary evolved through. 
When a BBH system evolves in isolation, it is expected to have component masses $m \lesssim \unit[65]{M_\odot}$ due to the effects of pair-instability supernovae~(e.g., \cite{HegerWoosley02, Fishbach17, TalbotThrane18}; see \cite{Belczynski2020} for a review of recent updates to this limit for various stellar populations). 
The co-evolution of the binary is thought to lead to component spins that are preferentially aligned with the orbital angular momentum~\citep{Kalogera2000, Campanelli06, Stevenson17dlk, TalbotThrane17}, and since compact binary orbits circularise through gravitational radiation at a faster rate than their separation reduces, any orbital eccentricity induced by the supernovae of the components becomes negligible by the time the gravitational-wave signal enters the observing band~\citep{Peters64, Hinder07}.\footnote{While Kozai-Lidov resonance~\citep{Kozai62, Lidov62} is predicted to lead to eccentric mergers and mis-aligned spins, the Kozai-Lidov field merger rate is thought to be small~\citep[e.g.,][]{Silsbee16, Antonini17, Fishbach17a, Rodriguez18jqu, triplespin}.} 
When a BBH system forms and merges dynamically, its properties can be detectably different from those of isolated mergers.
In the dense environments that support dynamical formation, repeated BH or stellar mergers can give rise to binaries in which one or both components have masses within the pulsational pair-instability mass gap~\citep[e.g.,][]{Gerosa17,2020arXiv200609744S, Kremer2020, Kimball2020b}.
Because the components do not co-evolve, their spins may have any orientation relative to each other~\citep{Rodriguez16}, and few-body interactions and/or gravitational-wave captures can give rise to mergers with non-negligible eccentricity close to merger~\citep[see, e.g.,][]{2014ApJ...784...71S, Rodriguez18b, Gondan18, Samsing18, Zevin18, Kyle_data_paper}.
In globular clusters, $\sim 5\%$ of all BBH mergers are expected to have significant eccentricity close to merger ($e \geq 0.1$ at \unit[10]{Hz})~\citep{Samsing17, SamsingDOrazio18, Rodriguez18a, Rodriguez18b}.

Globular clusters are observed in great quantities, both inside our Galaxy and beyond; there are $\approx 160$ known globular clusters in the Milky Way~\citep{GCDatabase}, $\approx 500$ in the neighbouring Andromeda Galaxy~\citep{Peacock10}, and $\sim 12000$ in supergiant elliptical galaxies like M87~\citep{Tamura06}. 
Despite their prolific nature, it is not known how globular clusters form. 
Globular clusters contain stars that are thought to be among some of the most ancient in their host galaxy~\citep[for example, the globular cluster Hpl contains some of the most ancient stars ($\gtrsim 12\,$Gyr) in the Milky Way;][]{Kerber2019}, making their formation difficult to observe with electromagnetic telescopes. 
To date, the primary method to constrain cluster ages is main-sequence fitting of colour-magnitude diagrams \cite[e.g.,][]{Gratton1997, Sarajedini2007, Vandenberg2013} with a small subset of cluster ages also determined from the white dwarf cooling sequence \citep[e.g.,][]{Hansen2013, Garcia-Berro2014}.
Typical globular cluster age measurements have uncertainties of order \order{\unit[1]{Gyr}}; see \cite{Forbes2015, Forbes2018} and references therein.
It is hoped that the James Webb Space Telescope (JWST)---due to be launched in October, 2021~\citep{JWSTLaunchDate}---will be able to constrain cluster ages to within $\unit[1]{Gyr}$~\citep[e.g.,][]{Correnti16}.

Measurements of globular cluster ages and metallicities suggest two different globular cluster sub-populations: very old globular clusters, which are observed to have a wide range of ages and metallicities; and younger globular clusters, which have metallicities that anti-correlate with their ages~\citep{Hansen2013, ForbesBridges2010, Vandenberg2013, Leaman2013, Forbes2015}.
Current theories of globular cluster formation fall into two main categories: (i) clusters formed as a byproduct of active star formation in galaxy discs~\citep[e.g.,][]{Elmegreen2010, Shapiro2010, Kruijssen2015} and (ii) clusters formed due to the collapse of dark matter halos during or before the epoch of reionisation~\citep[e.g.,][]{FallRees, Harley14, Ramirez-Ruiz2015, Trenti15, Taysun16, Ma20}. 
In category (i), the formation \new{probability} follows the observed star formation rate~\citep[SFR;][]{MadauDickinson}, peaking at $z \approx 2.5$~\citep{Forbes2015, Forbes2018}, while in category (ii) the formation \new{probability} peaks at $6 \lesssim z \lesssim 12$~\citep{Forbes2015, Trenti15}.

Constraining the primary formation epoch of globular clusters will answer long-established questions in astrophysics. 
If globular clusters predominantly form before $z \approx 6$, they may play a leading role in the reionisation of the Universe~\citep[e.g.,][]{Ma20}. 
On the other hand, if the globular cluster formation \new{probability curve} follows the SFR, and the majority of star formation takes place in such environments~\citep[e.g.,][]{LadaLada2003}, then detailed understanding of cluster formation histories may place critical constraints upon the overall SFR.
If we know the formation epoch of clusters, then we can adjust N-body simulations to more correctly reproduce clusters observed at $z = 0$, thereby enhancing our physical descriptions of cluster initial conditions.
Our understanding of the role that globular clusters play in the evolution of galaxies---for example, whether globular clusters are early galaxies~\citep{ElmegreenElmegreen}, failed galaxies~\citep{FallRees}, or galaxy remnants~\citep{Majewski2000}---can also be improved by observing globular clusters as they form and evolve. 

As detectors improve, gravitational waves will allow us to trace compact binary mergers throughout cosmic time~\citep[see, e.g.,][]{Vitale2019, Safarzadeh2019}.
In turn, this will allow us to use gravitational waves as probes of cluster formation and evolution.
The current generation of detectors can observe events out to redshifts $z \lesssim 1.5$---not far enough for globular cluster formation to be traced through our observations. 
In order to examine globular cluster formation, we must wait for third-generation gravitational-wave observatories such as the Einstein Telescope~\citep{ET2010} and Cosmic Explorer~\citep{CosmicExplorer}.
These observatories, proposed to begin taking data ca. 2035, will be able to detect BBH mergers with total mass of order \unit[\order{100}]{$M_\odot$} out to redshifts $z \approx 30$, and GW150914-like mergers out to $z \approx 50$~\citep{ThirdGenMetrics}.

In this paper, we demonstrate the power of gravitational-wave observations as probes of globular cluster formation and evolution. 
In Section~\ref{subsec:GC_birth_rate}, we motivate a Gaussian mixture model describing the globular cluster formation \new{probability} over cosmic time. 
We explain the metallicity-dependent merger time distribution used to convert this underlying globular cluster \new{formation probability} to the BBH merger \new{probability} in Section~\ref{subsec:BBH_merge_rate}.
In Section~\ref{sec:method}, we outline our population inference method.
We test our ability to recover the underlying globular cluster formation \new{probability} in Section~\ref{sec:injection_studies}, \new{obtaining} population inference results using simulated third-generation gravitational-wave observatory data \new{with realistic uncertainties}.
\new{For these simulations, we use only mergers that are massive and rapid-merging---signatures of dynamical formation---as ``snapshots'' of the clusters at creation.}
\new{We find that we can measure the formation epochs of globular clusters to \unit[0.02]---\unit[0.6]{Gyr} precision at $99\%$ confidence after one year of third-generation gravitational-wave observations---\new{comparable to} the forecasted accuracy of JWST~\citep[e.g.,][]{Correnti16}---unless cluster formation primarily occurs during reionisation, in which case the precision with which we can locate a secondary lower-redshift formation epoch is reduced to \unit[\order{10}]{Gyr}}.
\new{If we use all cluster mergers instead of just the small fraction that we consider to be identifiable as such, our constraints on cluster formation epochs tighten by up to an order of magnitude}.
In Section~\ref{sec:caveats}, we state the assumptions and caveats underlying our model.
We conclude in Section~\ref{sec:conculsions}.

\begin{figure*}
    \centering
    \includegraphics[width=\textwidth]{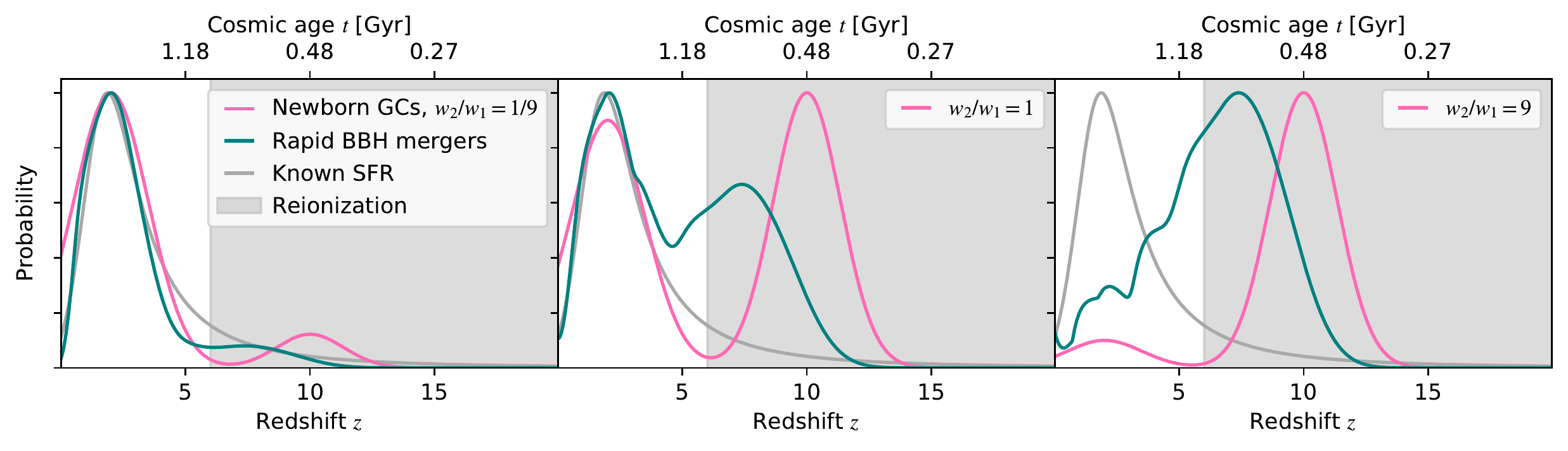}
    \caption{\new{The three cases we consider in this work are illustrated from left to right}: $\nicefrac{w_2}{w_1} = \nicefrac{1}{9}$, where SFR-driven globular cluster formation dominates clusters formed by $z=0$; $\nicefrac{w_2}{w_1} = 1$, where there is equal contribution to clusters formed by $z=0$ from both formation channels; and $\nicefrac{w_2}{w_1} = 9$, where reionisation-driven globular cluster formation dominates clusters formed by $z=0$. \new{The probability distributions of rapid-merging first-generation massive BBH mergers (where both components have $30 M_\odot \leq m \leq 40 M_\odot$) are plotted in teal, while the formation probability distributions of GCs in each model are plotted in pink}.}
    \label{fig:all-illustration}
\end{figure*}

\section{Model}
\label{sec:model}

In the following section, we describe our model, which combines theoretically-motivated globular cluster formation \new{probability} distributions (described in Section \ref{subsec:GC_birth_rate}) with simulated BBH merger distributions (described in Sections \ref{subsec:BBH_merge_rate} \new{and \ref{sec:subsets}}). 
We assume a flat $\Lambda$CDM Universe with $H_0=\unit[67.7]{km s^{-1} Mpc^{-1}}$ and $\Omega_0=0.307$~\citep{Planck2016}. 

\subsection{Globular Cluster Formation Probability Distribution}
\label{subsec:GC_birth_rate}

Our globular cluster formation \new{probability} \new{is modelled by a two-component Gaussian mixture model} in redshift. The first component represents SFR-driven globular cluster formation, with mean $\mu_1$, standard deviation $\sigma_1$ and weight $w_1$; the other represents reionisation-driven globular cluster formation, with mean $\mu_2$, standard deviation $\sigma_2$ and weight $w_2$. 


\begin{table}
\centering
\begin{tabular}{c||c|c} 
 & \multicolumn{2}{|c}{Value~} \\ 
\hline
Parameter & $z$ & $t$ [Gyr] \\
\hline
$\mu_1$ & $2.00$ & $3.30$ \\ 
$\sigma_1$ & $1.50$ & $2.32$ \\
$\mu_2$ & $10.00$ & $0.48$ \\
$\sigma_2$ & $1.35$ & $0.09$ \\
\hline \hline
$\nicefrac{w_2}{w_1}$ & \multicolumn{2}{|c}{$\nicefrac{1}{9},\quad 1,\quad 9$}\\
\hline
\end{tabular}
\caption{Parameter values chosen for our fiducial globular cluster \new{formation probability models}, used in the injection studies described in Section \ref{sec:injection_studies}. Each of the two Gaussian peaks in our model has mean $\mu_j$ and standard deviation $\sigma_j$, where $j=1$ refers to the SFR-driven peak and $j=2$ refers to the reionisation-driven peak. The ratio $\nicefrac{w_2}{w_1}$ determines the relative weight of the reionisation-driven peak against the SFR-driven peak. We vary $\nicefrac{w_2}{w_1}$ to test our ability to recover the underlying globular cluster \new{formation probability} in three different scenarios, between which the dominant formation mechanism of clusters varies.}
\label{tab:table_of_injected_parameters}
\end{table}

We simulate a fiducial globular cluster \new{formation probability} using specific parameter values shown in Table \ref{tab:table_of_injected_parameters}. For the injection sets, we set the mean and standard deviation of the SFR-driven peak in order to best represent the true shape of the SFR~\citep{MadauDickinson}. 
The mean of the reionisation-driven peak is motivated by the results of \cite{Trenti15}; see also \cite{Ramirez-Ruiz2015}.  

To investigate our ability to distinguish the preferred channel of globular cluster formation, we vary the weight ratio $\nicefrac{w_2}{w_1}$. 
We consider three cases: (i) that globular clusters are formed primarily as a byproduct of the SFR, and there is a small contribution formed during reionisation ($\nicefrac{w_2}{w_1} = \nicefrac{1}{9}$); (ii) that globular clusters are formed with equal probability during reionisation and through star formation ($\nicefrac{w_2}{w_1} = 1$); and (iii) that globular clusters are formed primarily during reionisation, with a small contribution forming in accordance with the SFR ($\nicefrac{w_2}{w_1} = 9$). \new{All three cases lead to a similar merger probability at $z=0$, so the scenarios cannot be distinguished by existing detectors. However, third-generation detectors will be able to constrain $\nicefrac{w_2}{w_1}$.} Our three globular cluster \new{formation probability functions are plotted in pink} in the three panels of Figure \ref{fig:all-illustration}. 

\subsection{Binary Black Hole Merger Probability Distribution}
\label{subsec:BBH_merge_rate}

\begin{figure*}
    \centering
    \includegraphics[width=0.75\textwidth]{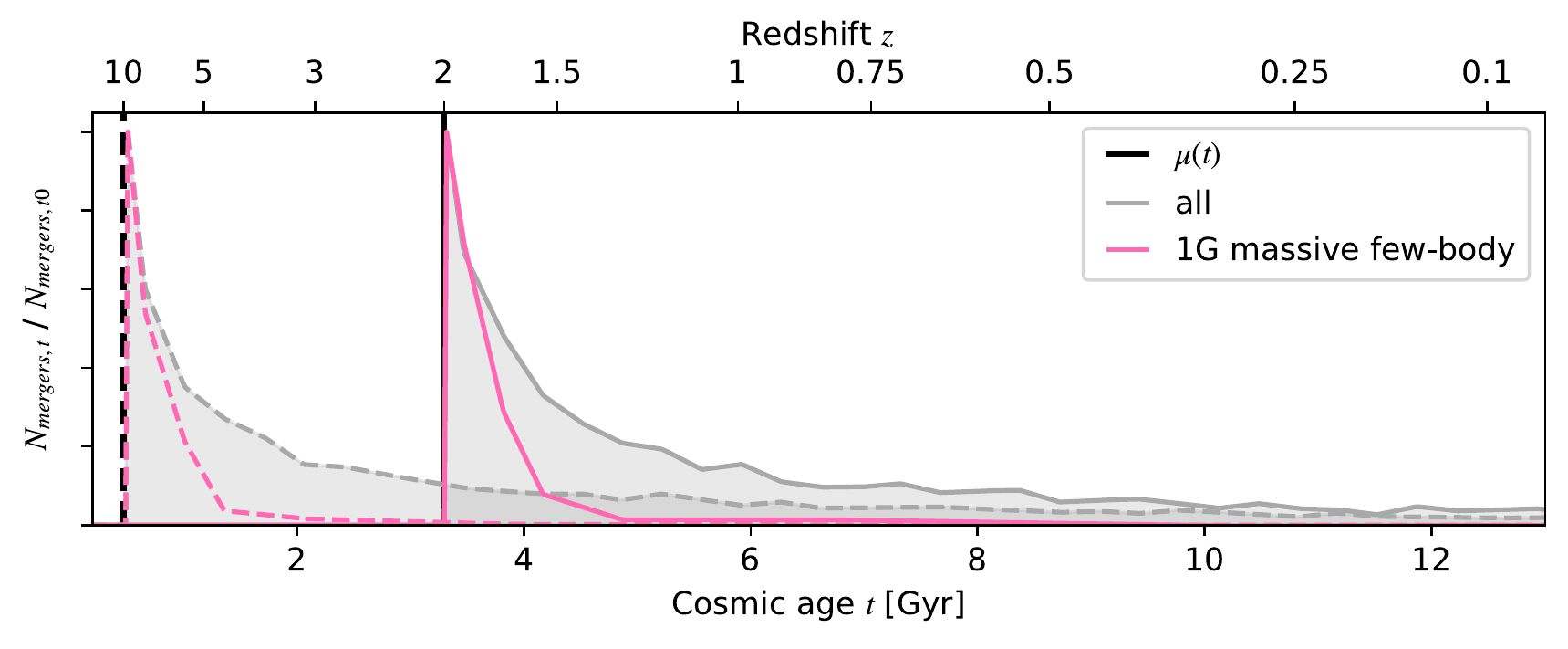}
    \caption{\new{Merger time distributions for BBH mergers from two clusters formed at $\mu_1$ ($z=10$; during the peak of the SFR) and $\mu_2$ ($z=2$; during reionisation), with distributions shown with solid and dashed curves respectively. The distributions of all mergers from each cluster after its \new{formation} are shown in grey. The distributions of just the massive first-generation ($\unit[35]{M_\odot} \leq m \leq \unit[40.5]{M_\odot}$; 1G) mergers formed through gravitational-wave capture during few-body interactions are shown in pink.}}
    \label{fig:two-clusters-delay-times}
\end{figure*}

In order to translate globular cluster formation probability into gravitational-wave observables, we calculate the distribution of BBH mergers in globular clusters.
To simulate globular cluster evolution, we use the \texttt{CMC Cluster Catalog} \citep{Kyle_data_paper}. These simulations were computed using \texttt{CMC} \citep{Joshi2000,Pattabiraman2013}, a H\'{e}non-type Monte Carlo code which includes various physical processes relevant to the dynamical formation of BH binaries including two-body relaxation, stellar and binary evolution \citep[computed using updated versions of \texttt{SSE} and \texttt{BSE};][]{Hurley2000,Hurley2002}, and direct integration of small-$N$ resonate encounters \citep{Fregeau2007} including post-Newtonian effects \citep{Rodriguez2018}. A number of parameters relevant to the long-term cluster evolution are varied within this set of simulations (namely the total cluster mass, initial virial radius, metallicity, and radial position within the Galactic potential), with values chosen to reflect the observed properties of the Milky Way globular clusters.
Altogether, this catalogue nearly completely covers the full parameter space of the Milky Way globular clusters and captures the formation of a variety of astrophysical objects such as gravitational-wave sources as well as X-ray binaries, pulsars, and blue stragglers. By implementing a cluster age distribution model from \citet{ElBadry2019}, \citet{Kyle_data_paper} estimated a BBH merger rate of roughly $20\,\rm{Gpc}^{-3}\,\rm{yr}^{-1}$ in the local Universe, consistent with previous theoretical work on the subject \citep[e.g.,][]{Rodriguez2016,Askar2017,RodriguezLoeb2018, Antonini2020} as well as with the observational rate inferred from the second LIGO/Virgo catalogue~\citep[][]{GWTC-2}.

Each newborn globular cluster in our model has a metallicity-dependent merger time distribution. 
The clusters in the \texttt{CMC Cluster Catalog} have one of three absolute metallicities: 0.0002, 0.002, and 0.02. We combine mergers from clusters with the same metallicity, then \new{sort the mergers into 100 time bins. We use a univariate spline to smoothly interpolate between the bin heights.} \new{We then perform a two-dimensional linear interpolation between these smoothed merger distributions of both cluster age and metallicity}. \new{To calculate the merger distribution for a cluster born at a certain redshift, we assume for simplicity that metallicity increases linearly with the age of the Universe. We assume a metallicity of $Z=0.0002$ at a redshift of 24 and a metallicity of $Z = 0.02$ at the present day}~\citep[see, e.g., Figure 6 of][for observationally-driven proposals for metallicity evolution over cosmic time]{Lilly02}.

\new{We convert the sum of merger time distributions from all clusters into a probability distribution in redshift, from which we draw our source population.
In Figure \ref{fig:all-illustration}, the merger probability distributions from each of the three globular cluster formation probability models are plotted with teal curves.
In Figure \ref{fig:two-clusters-delay-times}, we illustrate in grey the merger time distributions for two clusters: one formed at $z = 2$ and one formed at $z = 10$.}

\subsection{\new{Rapid mergers as cluster formation snapshots}}
\label{sec:subsets}
The merger time for a BBH formed dynamically in a stellar cluster is determined by three timescales: (i) the cluster \new{formation} time, (ii) the time required for BBH formation through dynamical encounters, and (iii) the gravitational-wave inspiral time from the time of last dynamical encounter to merger. The latter two timescales are sensitive to a variety of host cluster properties including total cluster mass, half-mass radius, and BH mass distribution \citep[e.g.,][]{Rodriguez16,Askar2017,Kyle_data_paper}. As a consequence, disentangling the cluster \new{formation} time distribution from the merger time distribution for a given list of dynamical BBH mergers may pose a challenge. This challenge may be circumvented by looking at specific classes of mergers known to have prompt merger times, $t_{merge} \lesssim \mathcal{O}(100\,\rm{Myr})$ \new{since cluster formation}. For these rapid mergers, the observed merger time distribution much more closely traces the underlying cluster \new{formation} time distribution.

Rapid mergers in globular clusters are expected to have two primary characteristics: high eccentricities and high masses. \new{During small-$N$ (`few-body') resonant encounters, pairs of BHs can form that merge rapidly, making them more likely to retain orbital eccentricity at detection.} As discussed in \citet{Samsing17} and \citet{Kyle_data_paper}, roughly $5-10\%$ of cluster mergers are expected to retain high eccentricity ($e \geq 0.1$) close to merger ($f_{\rm GW} = \unit[10]{Hz}$). 
These binaries can have gravitational-wave inspiral times as short as days \citep[e.g.,][]{Zevin18}; \new{this makes them ideal tracers of cluster formation, as they merge relatively quickly after the cluster forms and are more likely to retain the dynamically-induced eccentricity that can reveal their formation channel}. As a natural consequence of dynamical friction, the most massive BHs in a cluster are expected, on average, to be the first to form BBHs and the first to merge \citep[e.g.,][]{Morscher15}. 
Thus, BHs with masses near the assumed upper limit of the BH mass distribution \new{($40.5 M_{\odot}$ in the \texttt{CMC Cluster Catalog})} that merge through gravitational-wave capture encounters are ideal rapid merger candidates. \new{For the anaysis presented in Section \ref{sec:injection_studies}, we consider only globular cluster binaries that merge through resonant few-body encounters.}

\new{There is an inherent additional delay associated with second-generation BHs formed through previous BH mergers that remain bound to their host cluster.}
Although these will preferentially merge again quickly \new{(within a few $\unit[10]{Myr}$ of the previous merger)} due to their relatively high mass, we do not include second-generation mergers in this analysis.
\new{Here, we consider only those massive few-body mergers that are first-generation (1G), having \new{both component masses} above $35 M_{\odot}$ and below $40.5 M_{\odot}$}.
\new{We plot this distribution of mergers in pink in Figure \ref{fig:two-clusters-delay-times}.}
\new{Over the redshift range that we study, the fraction of 1G massive few-body mergers varies between 3\% and 6\% of all cluster mergers.}
\new{We construct the merger time distribution using only 1G massive few-body mergers, and draw only $5\%$ of the number of detections expected from the observing durations.}

\section{Method}
\label{sec:method}
We simulate \new{redshift posterior probability distributions for} a population of BBH mergers, and use the population inference framework to discern the injected distribution of the population. 
The likelihood for the data $\textbf{d}$ is 
\begin{equation}
\label{eq:poplikelihood}
    \mathcal{L}_{tot}(\mathbf{d} | \mathbf{\Lambda}) = \prod_{i}^{N} \frac{\mathcal{Z_\varnothing}(d_i)}{n_i}\sum_{k}^{n_i}\frac{\pi(\theta_i^k | \mathbf{\Lambda})}{\pi(\theta_i^k | \varnothing)}.
\end{equation}
In this equation, $\mathbf{\Lambda}$ is the set of parameters describing the population distribution, while $\theta_i^k$ are the parameters describing the $k$th posterior sample of event $i$ (in our case, $\theta_i$ is only one parameter -- \new{redshift}). 
Each event $i$ has $n_i$ posterior samples; there are a total of $N$ events. 
The sampling prior used for inference on data $d_i$ for event $i$ is $\pi(\theta_i | \varnothing)$, which is reweighted to obtain results for a population-based prior $\pi(\theta_i | \mathbf{\Lambda})$. 
The evidence obtained with the original sampling is $\mathcal{Z}_\varnothing(d_i)$.
In our case, the population prior $\pi(\theta_i | \mathbf{\Lambda})$ is the distribution described in Section~\ref{sec:model}.

\new{Distance measurement uncertainties are likely to be \order{10\%} for most binaries observed with third-generation detectors~\citep{Vitale2017, ZhaoWen2018} .
To model uncertainty of approximately this magnitude, we assume Gaussian likelihoods of width $\sigma_{z_i} = 0.1 z_i$. These likelihoods each have a mean $\mu_{z_i} = z_i + r_i$, where $r_i$ is a random offset drawn from a Gaussian of mean $\mu_{r_i} = 0$ and $\sigma_{r_i} = \sigma_{z_i}$. We produce a posterior curve by multiplying the likelihood by a uniform sampling prior. (This prior is divided out in the calculation of Eq. \ref{eq:poplikelihood}.) From this posterior curve, we draw 50 simulated posterior samples for each event.}

To execute our population analysis, we use the Bayesian inference library \texttt{bilby}~\citep{bilby, bilbyGWTC1}. 
We use uniform priors over all parameters. 
The prior covers the range $10^{-5} \leq z \leq 6$ for $\mu_1$, $0.5 \leq z \leq 6$ for both $\sigma$ values, and $6 \leq z \leq 20$ for $\mu_2$.
The prior on $\nicefrac{w_2}{w_1}$ ranges from $10^{-2}$ to $10$.

\new{The formation channel of a binary may be identified} using a method such as that developed in \cite{Kimball2020a}, in which a BBH merger's mass and spin measurements are used to calculate its probability of being a hierarchical merger in a globular cluster.
Similar methods may be extended to incorporate eccentricity measurements, which will be illuminating for globular clusters as we expect $\sim 5\%$ of globular cluster mergers to have eccentricity $e \geq 0.1$ at \unit[10]{Hz}~\citep[see, e.g.,][]{Samsing17}.
While precession is considered a hallmark of dynamical mergers, there have been relatively few events that have clear precession measurements~\citep{GWTC-1, GWTC-2}; however, both precession and anti-aligned spins can be measured at a population level, as demonstrated in~\citep{GWTC-2_RnP}.
With third-generation detectors, component spins and precession will be well measured~\citep[e.g.,][]{Vitale2018}. Using such measurements at both an individual and population level, it may be possible to estimate the sub-population of globular cluster mergers within a set of BBH mergers from a variety of formation channels.

\new{In this paper, we assume that mergers identified as cluster mergers have $0\%$ probability of having formed via a different channel.}
\new{Such definitive statements are unlikely to be made based on the parameters of detected binaries for the vast majority of sources} even if we allow for future improvements to our mechanisms for performing such identifications.
In the future, binaries that form in globular clusters but are kicked out before merging may still be indistinguishable from isolated mergers, and those that do merge inside the cluster are likely to have properties similar to those in other dynamical environments (e.g., AGN discs and galactic nuclei) or field triples undergoing Kozai-Lidov resonance. 
More complex future analyses \new{should} weight the samples from each event by the probability that each binary formed inside a globular cluster.
This is an additional complication that can be built upon the method presented here, and is left for future work.

\section{Injection studies}
\label{sec:injection_studies}

\begin{figure*}
     \centering
     \begin{subfigure}[b]{0.9\textwidth}
         \centering
        \includegraphics[width=\textwidth]{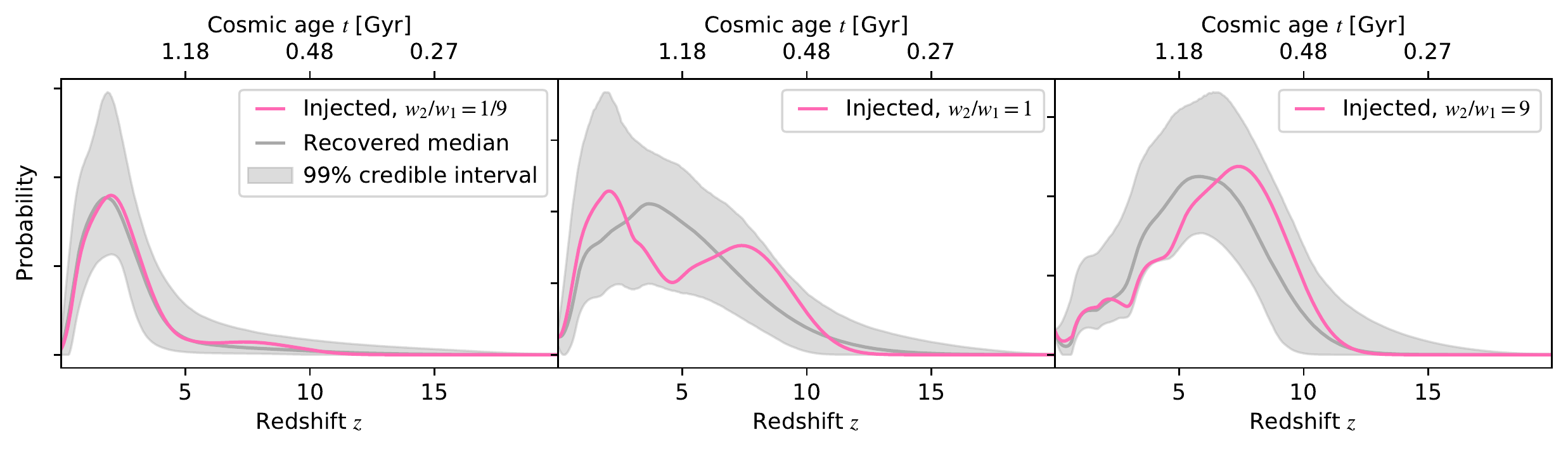}
        \caption{}
        \label{fig:day_recovered}
     \end{subfigure}
     \begin{subfigure}[b]{0.9\textwidth}
         \centering
    \includegraphics[width=\textwidth]{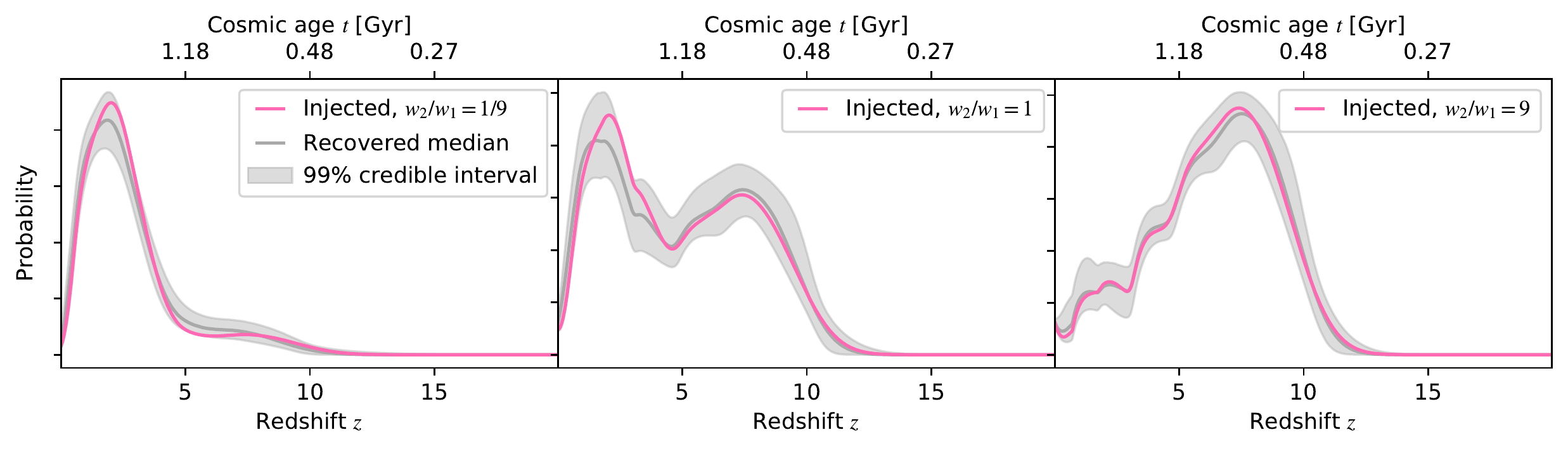}
    \caption{}
    \label{fig:month_recovered}
     \end{subfigure}
     \begin{subfigure}[b]{0.9\textwidth}
         \centering
    \includegraphics[width=\textwidth]{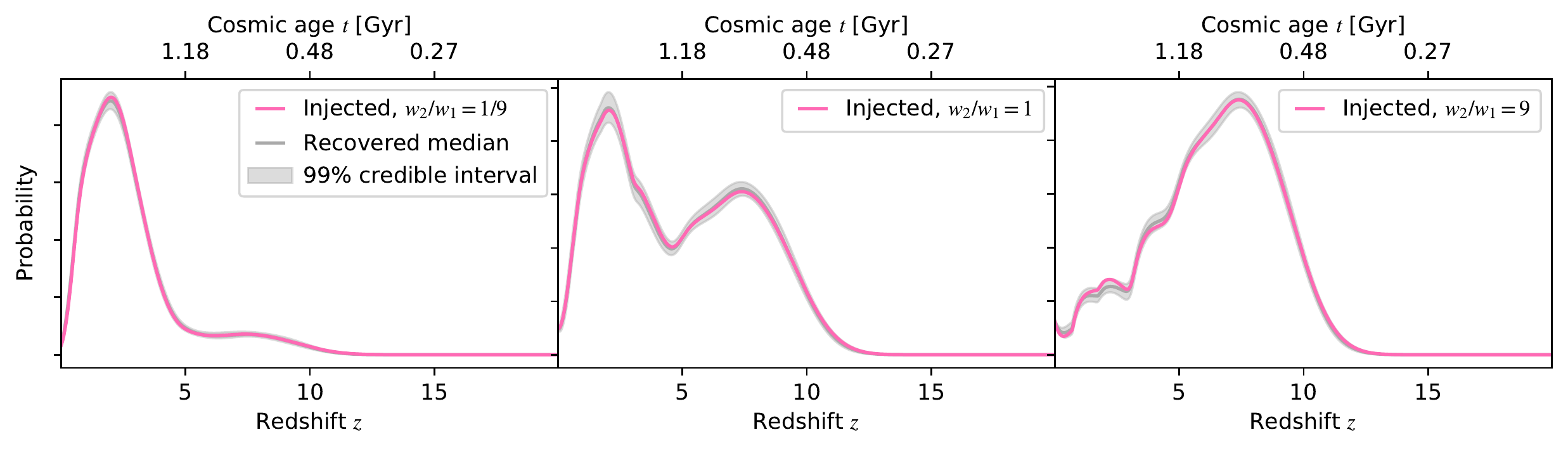}
    \caption{}
    \label{fig:year_recovered}
     \end{subfigure}
        \caption{Globular cluster formation rates \new{inferred from simulated third-generation detector observations of first-generation massive few-body mergers after (a) one day (using 25 1G massive few-body mergers from the total population of 500), (b) one month (using 500 of 10000) and (c) one year of detections (using 5000 of 100000).} Each panel represents an underlying model with a different weighting of the two components of the Gaussian mixture model that represents the globular cluster formation rate. \textit{Left}: SFR-driven peak dominates formation rate; \textit{Middle}: Each peak contributes equally to formation rate; \textit{Right}:  Reionisation-driven peak dominates formation rate.}
        \label{fig:recovered}
\end{figure*}

The Universal merger rate implied by current BBH merger observations is $\sim \unit[0.2]{min^{-1}}$ \citep{Abbott_GW170817_Stochastic_Implications}. 
For an all-seeing detector, this translates to a BBH signal detection count of $\mathcal{O}(500)$ per day~\citep{ET2020}. Third-generation detectors like CE and ET will be close to all-seeing, detecting GW150914-like events out to redshifts $z \approx 50$, and GW190521-like out to $z \approx 30$~\citep{ThirdGenMetrics}. \new{We use 500, 10000 and 100000 as order-of-magnitude estimates for the total number of GC mergers in one day, one month and one year, respectively. This is a reasonable approximation if GC mergers make up $\geq 10\%$ of all mergers in the Universe.} \new{We assume, based on the fraction of massive and quick-merging binaries observed in the cluster simulations described in Section \ref{sec:model}, that only $5\%$ of these observations can be confidently identified as cluster mergers. To simulate a month's worth of confidently-identified cluster mergers, for example, we use 500 events.}

We simulate data after one day, one month, and one year of observing, for three different models of the underlying globular cluster formation \new{probability}.
The first case we consider is one where globular clusters do not form efficiently during reionisation. In this case, the globular cluster formation \new{probability curve} closely follows the observed SFR. We set the weight ratio $\nicefrac{w_2}{w_1} = \nicefrac{1}{9}$, such that $90\%$ of all clusters form within the SFR peak. 
In the second case, we set the weight ratio $\nicefrac{w_2}{w_1} = 1$, which leads to a 50--50 split between clusters contributed from each peak.
The final case we consider is one where globular clusters primarily form during reionisation, such that $90\%$ of clusters are formed within the reionisation peak. We set $\nicefrac{w_2}{w_1} = 9$. 
For all three injected data sets, the remaining four parameters ($\mu_1, \mu_2, \sigma_1, \sigma_2$) are fixed to the values provided in Table \ref{tab:table_of_injected_parameters}.
We show the globular cluster formation \new{probability curve} and resulting \new{probability distribution of} observable BBH mergers for all three cases in Figure \ref{fig:all-illustration}.

\begin{table}
\centering
\begin{tabular}{|c|ccccc|} 
\multicolumn{1}{c|}{$\mu_1$} & & \multicolumn{4}{c|}{99\% CI \new{width} (Gyr) ~}  \\ 
\hline
&  $\nicefrac{w_2}{w_1}$   &                 & Day  & Month & Year        \\ 
\hline
& 9     &                       & \new{11.92} & \new{12.01}  & \new{10.49}                       \\
&  1     &                       & \new{11.99} & \new{9.87}  & \new{0.53}                       \\
&  \nicefrac{1}{9}  &            & \new{10.44} & \new{4.32}  & \new{0.39}             \\
\hline
\multicolumn{1}{c|}{$\mu_2$} & &  \\ 
\hline
&  9   &                         & \new{0.48} & \new{0.06}  & \new{0.02}                   \\
& 1     &                       & \new{0.74} & \new{0.08} & \new{0.02}                    \\
& \nicefrac{1}{9}    &         & \new{0.74} & \new{0.42}  & \new{0.05}         \\
\end{tabular}
\caption{Width of 99\% credible intervals (CIs) around $\mu_1$ \new{(top) and $\mu_2$ (bottom)} for each injection study described in Section \ref{sec:injection_studies}. \new{For these injection studies we use only first-generation massive few-body mergers, and include a 10\% uncertainty on source redshift. When the 10\% uncertainty is removed, the width of these uncertainty intervals does not meaningfully change.}}
\label{tab:table_of_intervals}
\end{table}

In Figure \ref{fig:recovered}, we plot the injected underlying globular cluster formation \new{probability curve} in pink, and compare it to the recovered median and 99\% confidence intervals in grey.
For all three observing periods, the injected distribution is within the 99\% confidence interval.
Probability distributions on the five populations parameters $\mu_1$, $\mu_2$, $\sigma_1$, $\sigma_2$ and $\nicefrac{w_2}{w_1}$ are provided as corner plots in Appendix \ref{appendix:posteriors}.
We state the widths of the 99\% confidence intervals around $\mu_1$ and $\mu_2$ for each study in Table \ref{tab:table_of_intervals}.

\new{We repeat the above injection studies for two additional scenarios.
In the first, we neglect any measurement uncertainty and assume that each source is represented by a delta function at its true value of $z$.  
}
\new{In this case, we see negligible change in the widths of the $99\%$ credible intervals around $\mu_1$ and $\mu_2$.}
\new{
In the second, we include redshift measurement uncertainty, but optimistically assume that all cluster mergers can be confidently identified, thereby allowing us to use 100\% of the mergers from the \texttt{CMC Cluster Catalog} to construct our model.
This leads to a reduction of up to an order of magnitude in the width of the 99\% credible interval around $\mu_1$ and $\mu_2$; for one day of observing all cluster mergers, the credible intervals are nearly identical to those seen for a month of observing only 1G massive few-body mergers.
}
\new{These measurements are more precise---despite the longer average merger timescale---because we are able to use 20 times as many events to probe the cluster formation rate.}
\new{The precision with which we can measure globular cluster formation epochs with third-generation observations, therefore, sensitively depends on the number of events that are confidently identified as globular cluster mergers.}

\section{Systematic error analysis and caveats}
\label{sec:caveats}

We make a number of simplifying assumptions and approximations in our analysis, allowing us to demonstrate a generic way to probe globular cluster formation using gravitational-wave detections. 
These are listed in this section, with the aim to reduce the number of assumptions we make in future work that builds upon this paper.

We approximate both epochs of globular cluster formation as simple Gaussians in redshift.
However, the true SFR determining the shape of the cluster formation \new{probability} does not follow a Gaussian, and the shape of the reionisation-driven cluster formation \new{probability} is not known.
More complex future extensions of this work may allow the shape of the Gaussians to vary, with the skewness of the Gaussian a model variable.
\new{We assume that the redshift posterior distributions are also Gaussian, but the true shape of the uncertainty distribution would vary depending on the noise in the data containing each signal}.

We set our injection studies in an optimistic future where \new{merger channels can be perfectly distinguished}.
\new{While we analyse the merger distribution of only those globular cluster BBH that are high-mass, rapid-merging and highly likely to be detectably eccentric (signatures of dynamical origin) for our primary results}, we still ignore any possibility of contamination from other dynamical formation channels that produce mergers with similar properties, such as mergers in AGN or Kozai-Lidov~\citep{Kozai62, Lidov62} triples in the field.

We also do not account for any sources redshifting out-of-band due to high masses or high eccentricities at high redshift.
We do not account for the disruption/creation of globular clusters during galaxy mergers; while the increased star formation of merging galaxies should be absorbed into the SFR peak of our models, the shape of the merger time distribution at a given epoch will differ if clusters are disrupted/created at that time due to galaxy mergers, even if the overall number of globular clusters remains the same.

We assume that metallicity increases linearly with the age of the Universe to obtain different merger time distributions for clusters born at different times, but do not consider a time-evolving initial mass/density function. 
Metallicity, mass and density may become globular cluster variables in future iterations of our model, allowing us to consider whether the physical properties of clusters can be inferred from observations of gravitational waves.

\section{Conclusions}
\label{sec:conculsions}
In this work, we show that observations from third-generation gravitational-wave detectors \new{will} allow us to \new{measure} the formation epochs of the population of globular clusters.
\new{Our primary results are obtained assuming that only first-generation massive few-body mergers can be confidently identified as globular cluster mergers.}
\new{If up to 50\% of clusters are born following the SFR, a cluster formation epoch at $z=2$ can be resolved to within less than \unit[1]{Gyr} precision after one year of observing; however, if the majority of clusters are born during reionisation, the time of this epoch will have an uncertainty \unit[\order{10}]{Gyr}.}
\new{For all scenarios, a cluster formation epoch at $z=10$ can be resolved to within less than \unit[1]{Gyr} precision after just one day of observing, and to within \unit[0.05]{Gyr} after a full year.}
\new{If we are able to confidently identify all globular cluster mergers as such, these uncertainty bands can decrease by up to an order of magnitude.}
With third-generation detectors Cosmic Explorer and Einstein Telescope due to commence observing in the 2030s~\citep{CE-2019, ET2020}, the question of how globular clusters formed may be answered with gravitational-wave observations within the next twenty years.
These results will be complementary to measurements from electromagnetic telescopes, such as JWST, which will be able to constrain the ages of individual clusters to within \unit[\order{1}]{Gyr}~\citep[e.g.,][]{Correnti16}.

Our method can be extended for use in multiple future projects.
For example, we could assign a population of mergers from multiple formation channels with a certain probability of being globular cluster mergers, and weight their contribution to the total merger distribution accordingly.
We could also use intrinsic binary parameters, e.g. mass and eccentricity, to infer properties of their host clusters, like their \new{densities at formation}.
The simple procedure outlined in this paper must be refined before application to real data.
In particular, future work should address the fact that many signals from the more massive and highly-eccentric sources may be redshifted out-of-band.
We leave potential extensions and improvements for future work.

\section*{Data Availability}

We use publicly available output from the globular cluster simulations of ~\cite{Kyle_data_paper}, and perform our analysis using publicly available Bayesian inference library \texttt{bilby}~\citep{bilby, bilbyGWTC1}.

\section*{Acknowledgements}

We thank Duncan Forbes and Michele Trenti for sharing with us their expertise in globular cluster formation scenarios. We also thank Michela Mapelli and Christopher Berry for their comments, which improved the manuscript, \new{and our anonymous reviewer, whose suggestions improved both the paper and the science it contains}. KK is supported by an NSF Astronomy and Astrophysics Postdoctoral Fellowship under award AST-2001751. PDL and ET are supported through Australian Research Council Future Fellowships FT160100112 and FT150100281, ARC Discovery Project DP180103155, and ARC Centre of Excellence CE170100004. JS is supported by the European Unions Horizon 2020 research and innovation programme under the Marie Sklodowska-Curie grant agreement No. 844629.

\bibliographystyle{mnras}
\bibliography{main} 

\appendix

\newpage 
\section{Posterior probability distributions for population parameters}
\label{appendix:posteriors}

In this section, we present posterior probability distributions for our five population parameters in cosmic time\new{, recovered using the merger-time distributions of 1G massive few-body mergers}. Posterior distributions are plotted in grey, with the injected values indicated with pink lines, and the shading gradients on the two-dimensional posteriors indicate levels of $1\sigma$, $2\sigma$ and $3\sigma$ credibility, while the dashed grey error bars around the median recovered values show 99\% confidence intervals.

\onecolumn

\newpage
\subsection{Results after one day of observing}

\begin{figure}
     \centering
     \begin{subfigure}[b]{0.49\textwidth}
         \centering
        \includegraphics[width=\textwidth]{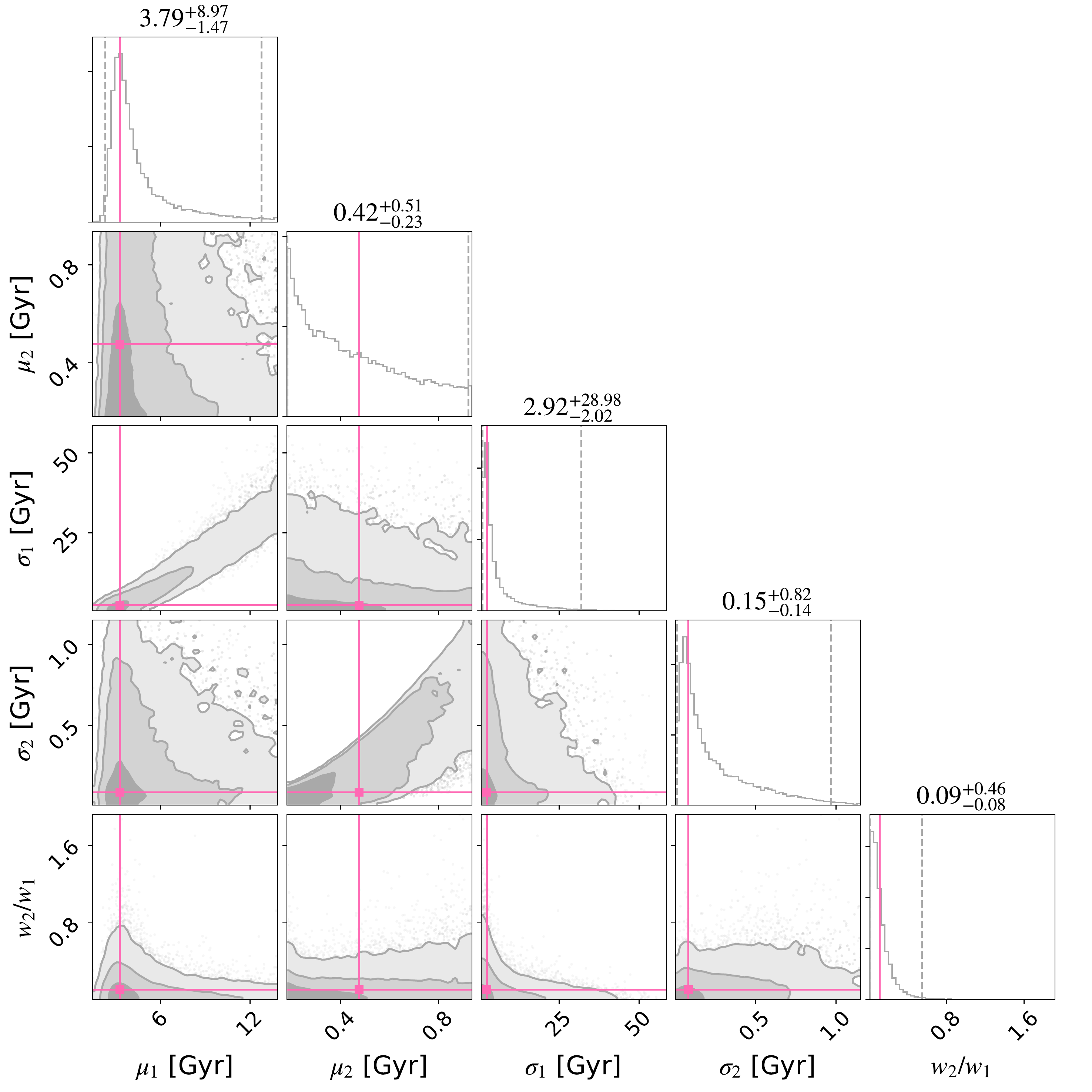}
        \caption{}
        \label{fig:90-10_Day_recovered}
     \end{subfigure}
     \begin{subfigure}[b]{0.49\textwidth}
         \centering
    \includegraphics[width=\textwidth]{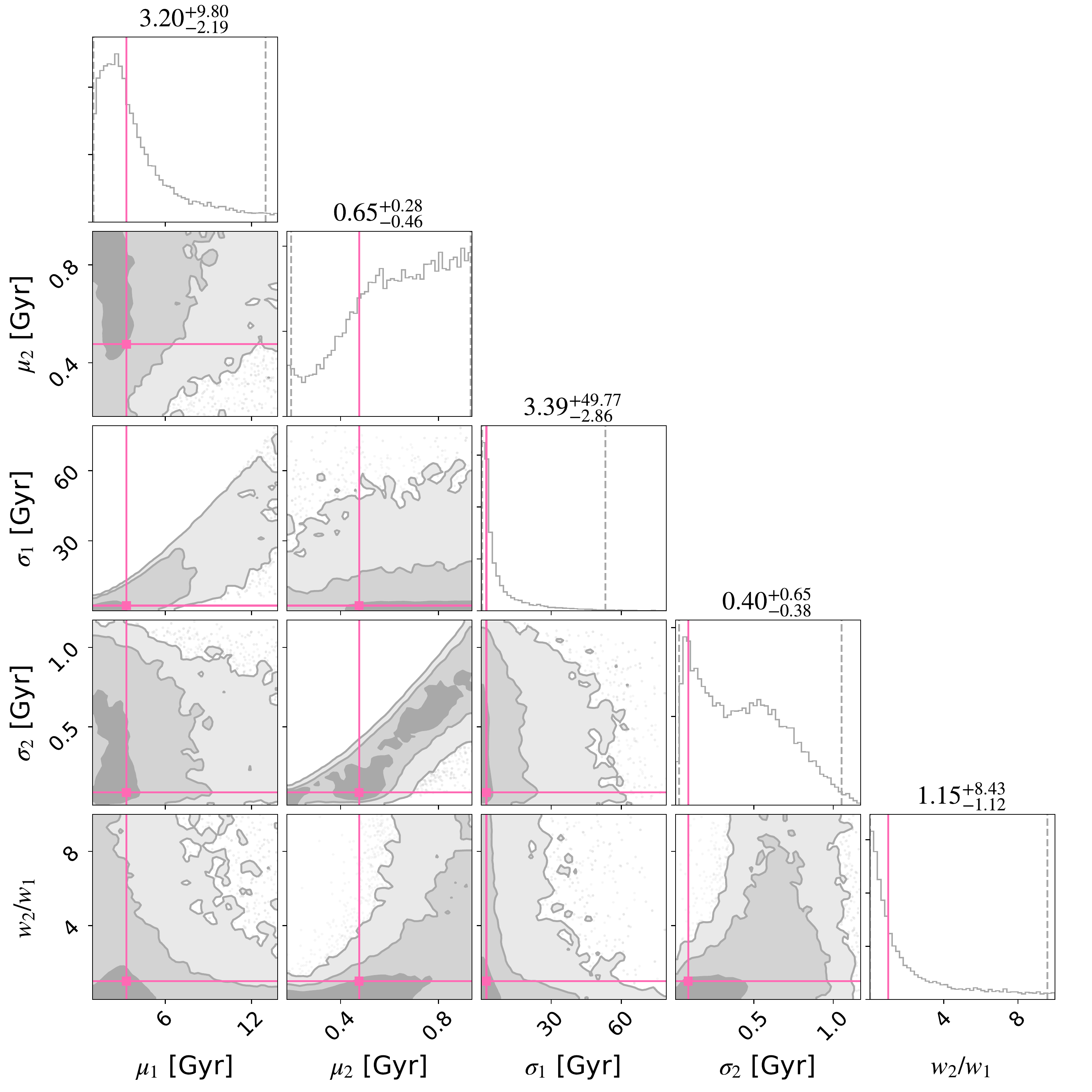}
    \caption{}
    \label{fig:50-50_Day_recovered}
     \end{subfigure}
     \begin{subfigure}[b]{0.49\textwidth}
         \centering
    \includegraphics[width=\textwidth]{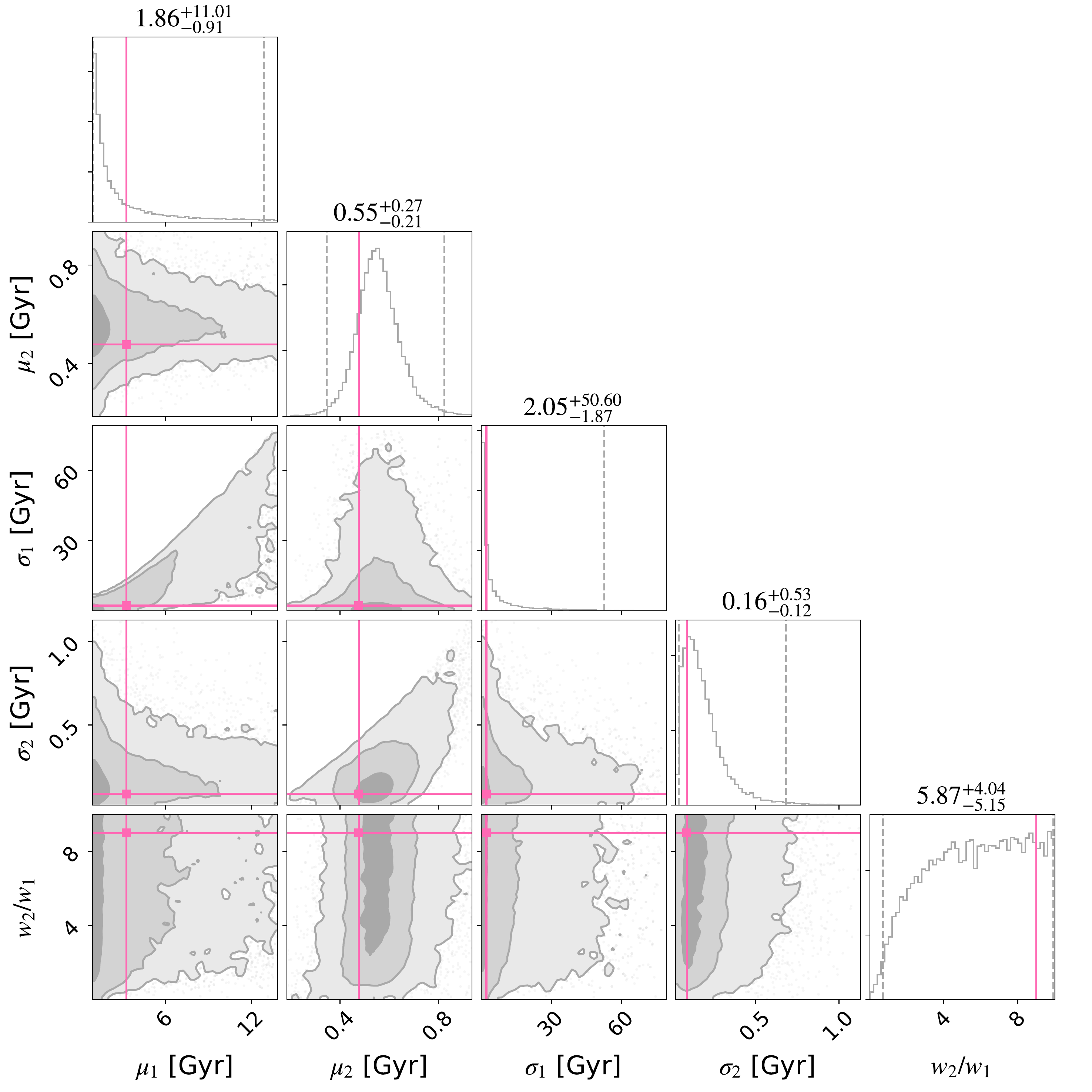}
    \caption{}
    \label{fig:10-90_Day_recovered}
     \end{subfigure}
        \caption{\new{Posterior probability distributions on our five population parameters after one day of simulated third-generation detector observations of massive few-body mergers from globular clusters (25 events). Results are shown for three variations on our two-component Gaussian mixture model: (a) $\nicefrac{w_2}{w_1} = \nicefrac{1}{9}$, (b) $\nicefrac{w_2}{w_1} = 1$, and (c) $\nicefrac{w_2}{w_1} = 9$.}}
        \label{fig:one-day}
\end{figure}

\new{In Figure \ref{fig:one-day} we present the one- and two-dimensional posterior probability distributions over each population parameter resolved after one day of observations with third-generation detectors.}

\newpage
\subsection{Results after one month of observing}

\begin{figure}
     \centering
     \begin{subfigure}[b]{0.49\textwidth}
         \centering
        \includegraphics[width=\textwidth]{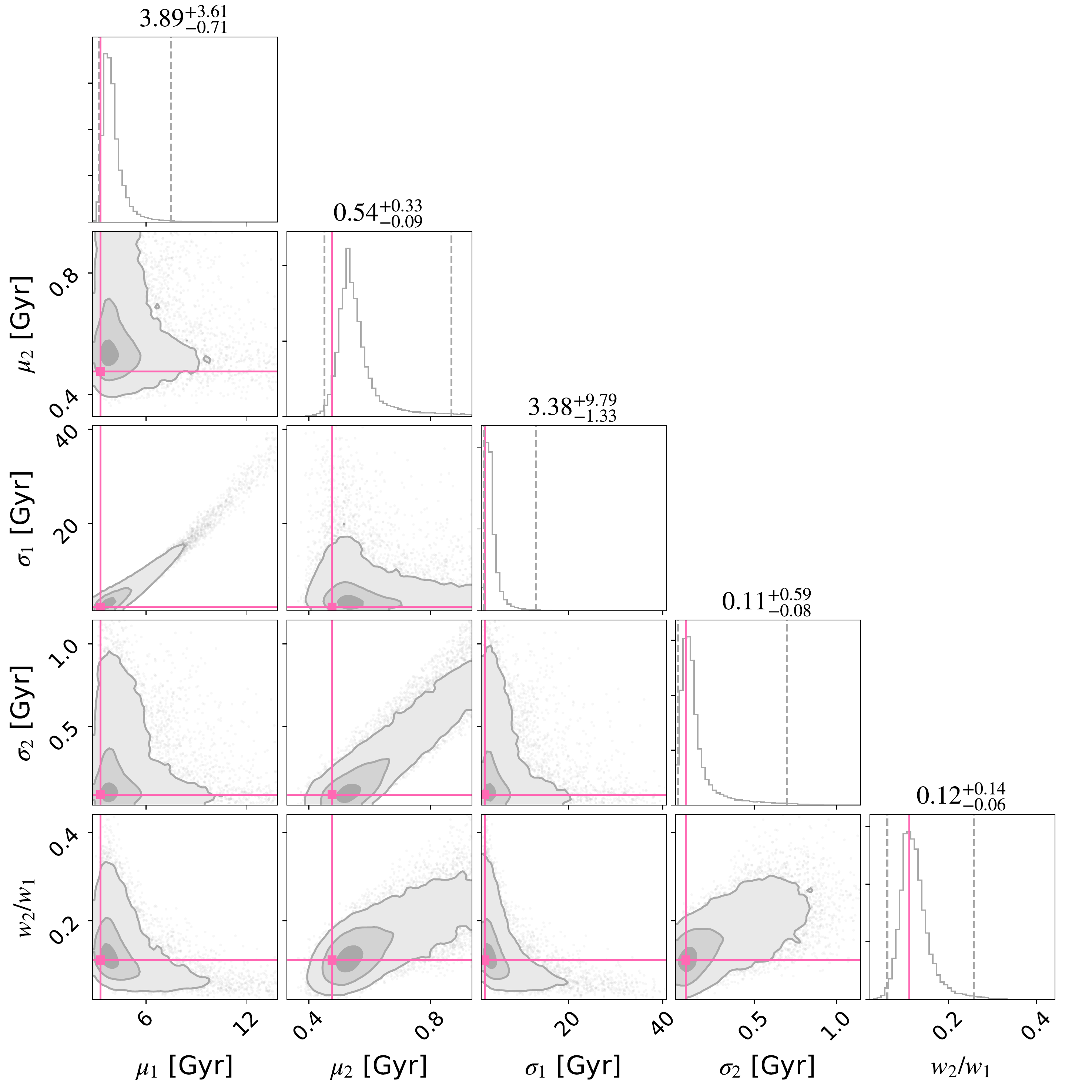}
        \caption{}
        \label{fig:90-10_Month_recovered}
     \end{subfigure}
     \begin{subfigure}[b]{0.49\textwidth}
         \centering
        \includegraphics[width=\textwidth]{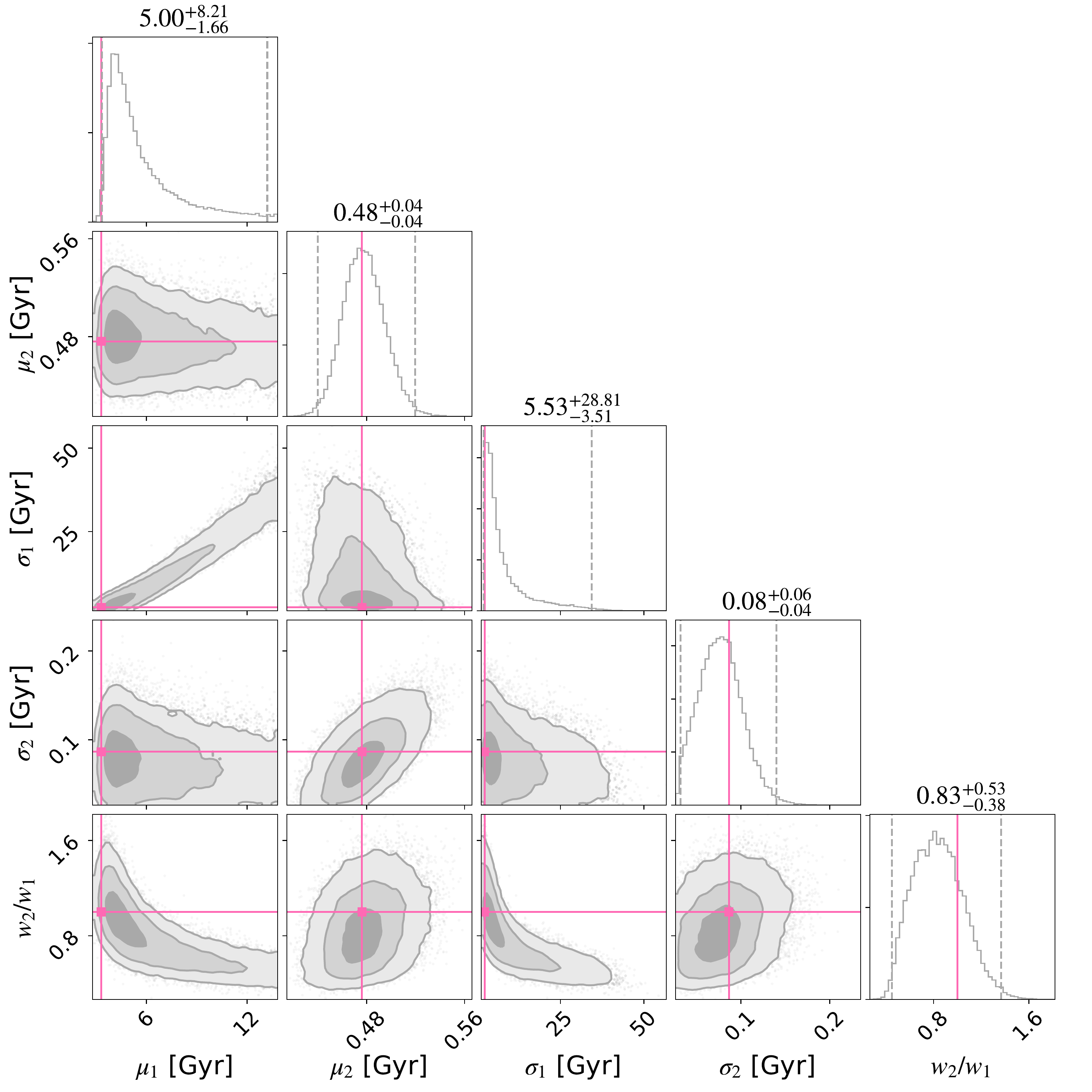}
        \caption{}
        \label{fig:50-50_Month_recovered}
     \end{subfigure}
     \begin{subfigure}[b]{0.49\textwidth}
         \centering
        \includegraphics[width=\textwidth]{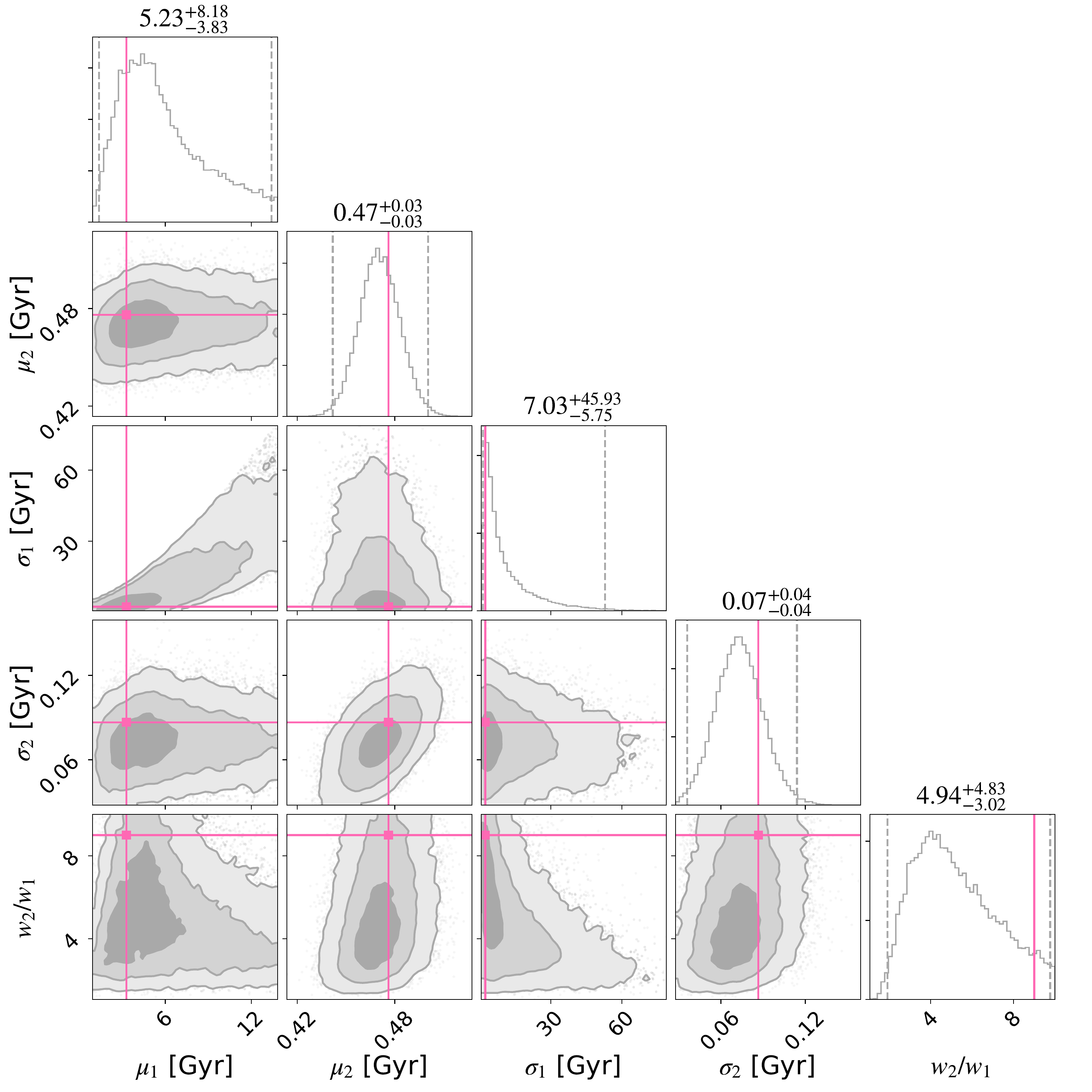}
        \caption{}
        \label{fig:10-90_Month_recovered}
     \end{subfigure}
        \caption{\new{Posterior probability distributions on our five population parameters after one month of simulated third-generation detector observations of massive few-body mergers from globular clusters (500 events). Results are shown for three variations on our two-component Gaussian mixture model: (a) $\nicefrac{w_2}{w_1} = \nicefrac{1}{9}$, (b) $\nicefrac{w_2}{w_1} = 1$, and (c) $\nicefrac{w_2}{w_1} = 9$.}}
        \label{fig:one-month}
\end{figure}

\new{In Figure \ref{fig:one-month} we present the one- and two-dimensional posterior probability distributions over each population parameter resolved after one day of observations with third-generation detectors.}

\newpage
\subsection{Results after one year of observing}
\begin{figure}
     \centering
     \begin{subfigure}[b]{0.49\textwidth}
         \centering
        \includegraphics[width=\textwidth]{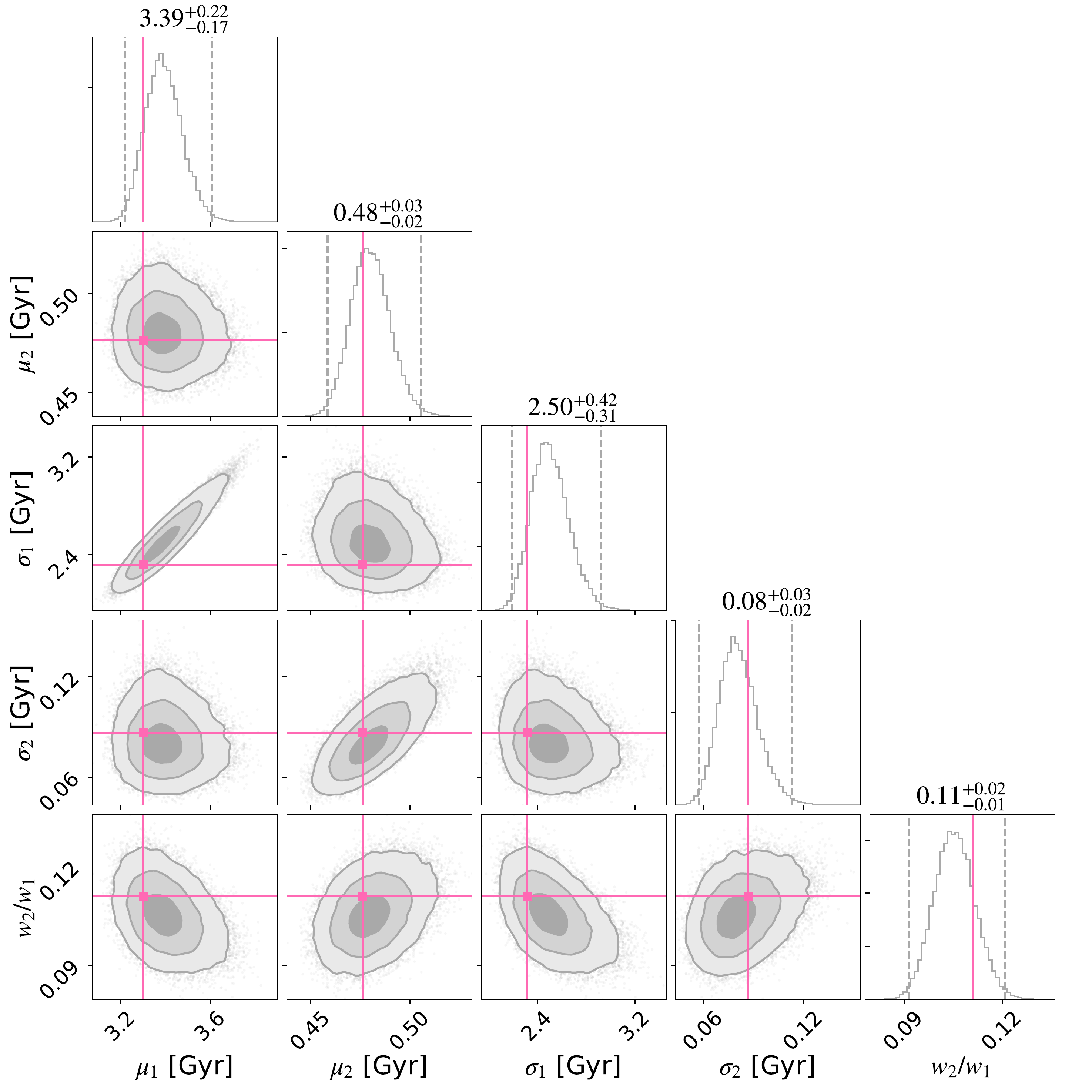}
        \caption{}
        \label{fig:90-10_Year_recovered}
     \end{subfigure}
     \begin{subfigure}[b]{0.49\textwidth}
         \centering
        \includegraphics[width=\textwidth]{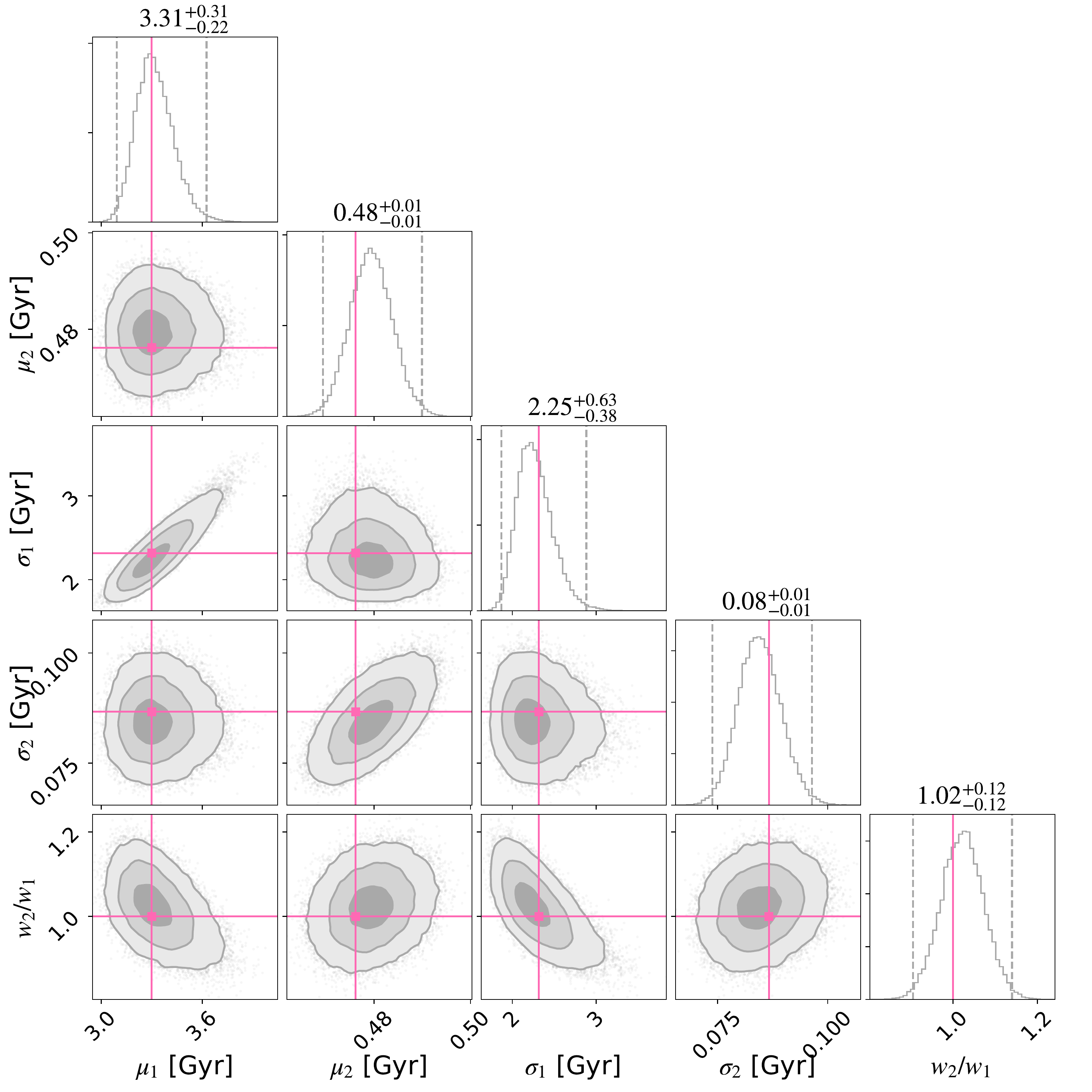}
        \caption{}
        \label{fig:50-50_Year_recovered}
     \end{subfigure}
     \begin{subfigure}[b]{0.49\textwidth}
         \centering
        \includegraphics[width=\textwidth]{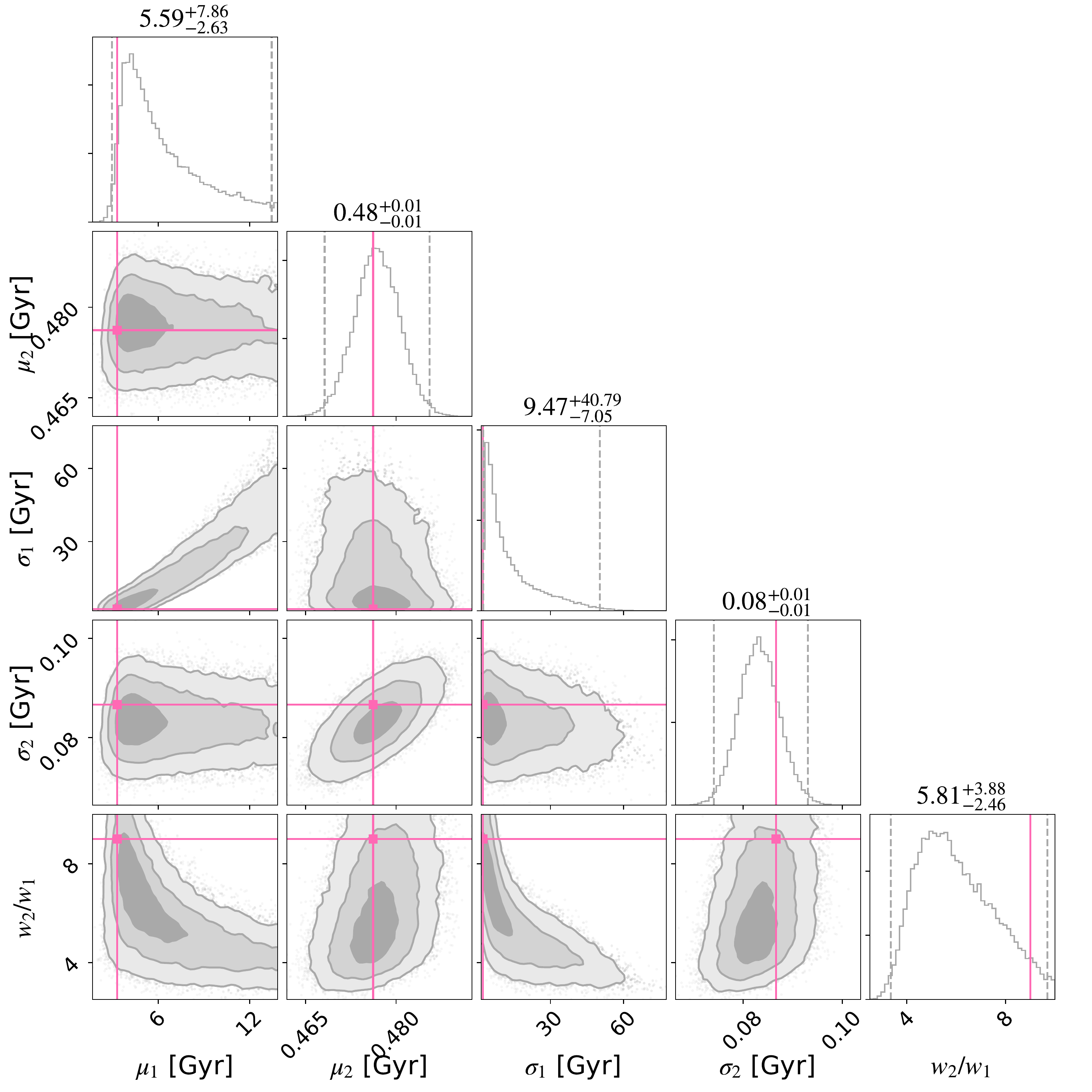}
        \caption{}
        \label{fig:10-90_Year_recovered}
     \end{subfigure}
        \caption{\new{Posterior probability distributions on our five population parameters after one year of simulated third-generation detector observations of massive few-body mergers from globular clusters (5000 events). Results are shown for three variations on our two-component Gaussian mixture model: (a) $\nicefrac{w_2}{w_1} = \nicefrac{1}{9}$, (b) $\nicefrac{w_2}{w_1} = 1$, and (c) $\nicefrac{w_2}{w_1} = 9$.}}
        \label{fig:one-year}
\end{figure}

\new{In Figure \ref{fig:one-year} we present the one- and two-dimensional posterior probability distributions over each population parameter resolved after one day of observations with third-generation detectors.}

\bsp	
\label{lastpage}
\end{document}